\documentclass[aps,prd,preprint,superscriptaddress]{revtex4-1}

\usepackage{amssymb,bm,graphicx,natbib}
\usepackage{hyperref,dcolumn}

\begin{document}


\title{\textbf{Mass of Galactic Lenses in Modified Newtonian Dynamics}}



\author{Mu-Chen Chiu}
\affiliation{Institute of Astronomy, National Central University,
           Jhongli, Taiwan 320, R.O.C.}
\affiliation{Scottish University Physics Alliance, Institute for
Astronomy, the Royal Observatory
 , University of Edinburgh, Blackford Hill, Edinburgh,EH9 3HJ
,UK}

\author{Chung-Ming Ko}
\affiliation{Institute of Astronomy, National Central University,
          Jhongli, Taiwan 320, R.O.C.}
\affiliation{Department of Physics, National Central University,
          Jhongli, Taiwan 320, R.O.C.}
\affiliation{Center for Complex Systems, National Central
University, Jhongli, Taiwan 320, R.O.C.} \email[corresponding
author:]{cmko@astro.ncu.edu.tw}

\author{Yong Tian}
\affiliation{Department of Physics, National Central University,
Jhongli, Taiwan 320, R.O.C.} \email{yonngtian@gmail.com}

\author{HongSheng Zhao}
\affiliation{Scottish University Physics Alliance, University of
St. Andrews, KY16 9SS,UK} \email[corresponding
author:]{hz4@st-andrews.ac.uk}


\date{\today}

\begin{abstract}
Using strong lensing data Milgrom's MOdified Newtonian Dynamics
(MOND) or its covariant TeVeS (Tensor-Vector-Scalar Theory) is
being examined here as an alternative to the conventional
$\Lambda$CDM paradigm. We examine 10 double-image gravitational
lensing systems, in which the lens masses have been estimated by
stellar population synthesis models. While mild deviations exist,
we do not find out that strong cases for outliers to the TeVeS
theory.
\end{abstract}



\maketitle


\section{\label{sec:intro}Introduction}
The concordant $\Lambda$CDM paradigm \cite[see e.g.,][]{teg06} has
been accepted as a successful framework to understand the
evolution of the universe. However, in this standard model there
are two unexplained dark sectors, dubbed dark matter and dark
energy, still need more fundamental understandings. One of various
endeavors on this explanation is the approach of modified gravity,
including Bekenstein's TeVeS \cite{TeVeS}. Unlike most theories of
modified gravity that are proposed for explaining dark energy,
TeVeS was originally proposed to explain dark matter instead.
Indeed, TeVeS was built up by Bekenstein for a viable relativistic
version of Milgrom's MOdified Newtonian Dynamics (MOND) \cite{M1}.
Other many recent theories which recovers MOND in it's limit
include the bimetric theory of ~\cite{M2009} and the dark fluid
theory of~\cite{zl10}.


In MOND, the demand of the exotic dark matter is replaced by the
modification of Newton's second law:
\begin{equation}
    \tilde\mu(|\mathbf{a}|/\mathfrak{a}_0)\mathbf{a} =
  -\mbox{\boldmath{$\nabla$}}\Phi_{\rm N},
    \label{MOND}
\end{equation}
with $\mathfrak{a}_0\approx 1.2\times10^{10}$ m s$^{-2}$.
$\tilde\mu(x)$ is called interpolation function in literature
sometimes. $\tilde\mu(x)\approx x$ for $x\ll 1$ (the deep MOND
regime), and $\tilde\mu(x)\rightarrow 1$ when $x\rightarrow 1$
(the Newtonian regime). Here $x=|\mathbf{a}|/\mathfrak{a}_0$, the
ratio of the acceleration to $\mathfrak{a}_0$, is a measure of
modified gravity.
MOND --in its relativistic version TeVeS \cite{TeVeS}--is not only
able to explain CMB successfully with 2 eV neutrino mass
\cite{SMFB06,dl06} but also even more successful than the
$\Lambda$CDM paradigm on the dynamics of spiral galaxies
\cite{SandersMcGaugh,Fam07}.

Unlike cold dark matter, however, massive neutrino cannot
aggregate at galactic scales and below, so any evidence for the
demand of dark matter at these scales is devastating to MOND.
Beside dynamical analysis on rotation curve (of spiral galaxies),
gravitational lensing offers us another way to investigate MOND
and TeVeS at these scales. Indeed, the requirement of dark matter
in the bullet cluster makes gravitational lensing even more
important \cite{clowe06}.

The earliest works about gravitational lensing in MOND were
studied by ~\cite{Qin95} and ~\cite{mort1}. Although their works
were performed even before the appearance of a viable general
covariant MOND, and their calculations on the angle of deflection
were artificially forced into the deep MOND regime when the
acceleration is less than $\mathfrak{a}_0$, i.e., $\tilde\mu(x)=
x$ for $x\leq 1$ (not $x\ll 1$), ~\cite{Qin95} and ~\cite{mort1}
could show a good guess on how to calculate lensing in MOND. After
the advent of TeVeS, a first calculation on light bending in TeVeS
was done by Bekenstein himself, and was followed up by
\cite{ckt06}. Based on a point mass model, \cite{ckt06} apply the
formalism of gravitational lesing in TeVeS to the theoretical
discussion on angle of deflection, magnification, as well as time
delay.

On the other hand, taking from a more phenomenal approach,
~\cite{zhaoeal06} showed that TeVeS is consistent with a sample of
double-image lenses in CASTLES catalogue by modeling lenses as
Hernquist model. ~\cite{Feix08} studied the effects of asymmetric
systems on gravitational lensing. ~\cite{shan08} then applied
non-spherical model to investigate strong lensing in TeVeS, and
found 10 out of 15 systems are consistent with TeVeS. All of other
5 systems are found to reside in or close to clusters of galaxy,
where external field could have significant influence.
Furthermore, the effect of large filaments on gravitational
lensing was studied by ~\cite{xu08}. They argued that filamentary
structures might have complex but significant contribution to the
system such as bullet cluster, so the need of dark matter in
bullet cluster might be spared again in MOND. However, all of
these work above are non-relativistic approximation of TeVeS. In a
conference paper, ~\cite{ckt08} also showed the lens data from
CASTLES and SLACS catalog are consistent with TeVeS, but they did
not show the details.

The first effort to calculate gravitational lesing from first
principle was given by ~\cite{mav09}. Along with their other works
~\cite{Ferr08,Ferr09}
, they showed that TeVeS might still need dark matter to explain
the lensing systems from CASTLES, and is lack of consistent
results between dynamical and gravitational lensing
analysis~\citep{Ferr08,Ferr09}.

Due to the importance of studying gravitational lesning in TeVeS,
and the inconsistency of the results on this field, in this paper,
we are going to apply the relativistic approach developed by
~\citet{TeVeS} and ~\citet{ckt06}, and the Hernquist model (used
for TeVeS firstly by ~\citet{zhaoeal06}) to the 10 double-image
systems from a total of 18 objects studied in~\citet{Ferr05}. In
this paper we will also clarify the underlying assumptions of
lensing calculation in TeVeS. The structure of the paper is
organized as below. In Sec.~\ref{formalism} we will briefly
outline the formalism of gravitational lensing in TeVeS. We will
also discuss how gravitaional lensing in TeVeS will depend on
different choice of $\tilde\mu(|\mathbf{a}|/\mathfrak{a}_0)$, and
its application to double-image systems. We then will give our
result in Sec.~\ref{deflectionAngle}, and discussion in
Sec.~\ref{discussion}.
\section{Formalism of gravitational lensing in TeVeS}\label{formalism}


\subsection{Lensing in TeVeS}
While considering strong lensing in TeVeS, we take the assumption
that in the weak field limit, the physical metric of a static
spherically symmetric system can be expressed in the isotropic
form (c=1):
\begin{equation}\label{weakspherical}
  \tilde{g}_{\alpha\beta}\, dx^\alpha dx^\beta
  = -(1+2\Phi)dt^2+(1-2\Phi)[d\varrho^2+\varrho^2(d\theta^2+\sin^2\theta d\varphi^2)]\,,
\end{equation}
where $\Phi=\Xi\Phi_{N}+\phi$ and $\Xi\equiv e^{-2
\phi_{c}}(1+K/2)^{-1}$ with $\phi_{c}$ as the asymptotic boundary
value of $\phi$ and $K<10^{-3}$ from the constrain of PPN
parameters
\citep{TeVeS}. Since in TeVeS all kinds of matter are coupled to
physical metric rather than Einstein's metric, the connection in
the geodesic equation,
\begin{equation}\label{geodesic}
  {{d^2x^{\mu}\over dp^2}+\Gamma^{\mu}_{\nu\lambda}{dx^{\nu}\over dp}
  {dx^{\lambda}\over dp}}=0\,,
\end{equation}
has to be constructed from physical metric. Here $p$ is some
proper affine parameter.
In the case of light bending by a static spherically symmetric
lens,
we can apply Eq.(\ref{weakspherical}) to the geodesic equation,
Eq.(\ref{geodesic}).
Consider a photon propagates on a null geodesic, and moves in the
equatorial plane, i.e., $\theta=\pi/2$, the geodesic equation or
equations of motion can be written as
\begin{eqnarray}
  (1+2\Phi) \dot t=E\,, \label{1stcmotion}\\
  (1-2\Phi) \varrho^2\dot\varphi=L\,, \label{2ndcmotion}\\
  (1-2\Phi) \dot{\varrho}^{2}+(1+2\Phi)\varrho^{-2}{L}^2-(1-2\Phi) E^2=0\,, \label{3rdcmotion}
\end{eqnarray}
where over dot denotes the derivative with respect to $p$. Recall
the fact that at the closest approach $\dot{\varrho}=0$,
$\varrho=\varrho_{0}$, we can express the ratio of the angular
momentum $L$ to energy $E$ as
\begin{equation}\label{impact}
  \ b^2 \equiv {L^2 \over E^2}={\varrho_{0}^2(1-2\Phi_{0})\over(1+2\Phi_{0})}\,,
\end{equation}
where $\Phi_{0}\equiv\Phi(\varrho_{0})$. Here $b$ is called the
impact parameter. Combining Eqs.~(\ref{2ndcmotion})-(\ref{impact})
gives the orbit of the photon,
\begin{equation}\label{orbit}
  -(1-4\Phi)+(1-4\Phi_0)\left(\varrho_0\over\varrho\right)^2
  \left[{1\over\varrho^2}\left(d\varrho\over d\varphi\right)^2+1\right]=0\,,
\end{equation}
of which the solution in quadrature is
\begin{equation}\label{angle1}
  \varphi=\int^\varrho{\left\{\left({\varrho\over\varrho_{0}}\right)^2
  \left[1-4(\Phi-\Phi_{0})\right]-1\right\}^{-1/2}{d\varrho \over \varrho}}\,.
\end{equation}
If we take the Taylor expansion of Eq.~(\ref{angle1}) to the first
order of $\Phi$, we get
\begin{equation}\label{angle1-2}
  \varphi \simeq \int^{\varrho}_{\varrho_0}
  {2\varrho_{0}{d\varrho}\over \varrho{\left(\varrho^2-\varrho_{0}^2\right)^{1/2}}}
  +\varrho_{0}\int^{\varrho}_{\varrho_0}
  {{4\varrho(\Phi-\Phi_{0})}\over{\left(\varrho^2-\varrho_{0}^2\right)^{3/2}}}{d\varrho}\,.
\end{equation}
The first term is the orbit without gravity, i.e., a straight line
($\varphi\rightarrow\pi$ as $\varrho\rightarrow\infty$). The
second term is the angle of the deviation due to the gravity.
Hence, to first order of $\Phi$, the deflection angle is
\begin{equation}\label{angle3}
  \Delta\varphi=\varrho_{0} \int^{\varrho}_{\varrho_0}
  {{4\varrho(\Phi-\Phi_{0})} \over {\left(\varrho^2-\varrho_{0}^2\right)^{3/2}}}{d\varrho}
  =\varrho_{0} \int^{\varrho}_{\varrho_0}
  {4|\bm\nabla\Phi|\over (\varrho^2-\varrho_{0}^2)^{1/2}}\ d\varrho
  -{{4\varrho_{0}\Phi}\over{({\varrho^2-\varrho_{0}^2})^{1/2}}}\,,
\end{equation}
where we have made use of integration by parts. Since for very
large $\varrho$, $\Phi$ behaves as $\ln\varrho$, it is legitimate
to ignore the second term of Eq.(\ref{angle3}) for strong lensing
systems~\citep{ckt06}. We then have the kernel equation of strong
gravitational lensing in TeVeS (up to first order of
$\bm\nabla\Phi$),
\begin{equation}\label{angle}
  \Delta\varphi=4\varrho_{0} \int^{\infty}_{\varrho_0}
  {|\bm\nabla\Phi|\over (\varrho^2-\varrho_{0}^2)^{1/2}}\ d\varrho\,.
\end{equation}
Recall that $\varrho_{0}$ is the distance of the closest approach,
and $\nabla\Phi$ is the MONDian gravity.


\subsection{Gravity in TeVeS}
In last subsection we have shown that, under the assumption of the
validity of Eq.(\ref{weakspherical}), the difference between the
angle of deflection in GR and that in TeVeS only arises from
$\nabla\Phi$. In this subsection, we are going to discuss the
relation between Newtonian gravity, $\nabla\Phi_N$, and MONDian
gravity, $\nabla\Phi$, and its application to strong lensing.

It has been shown in quasi-static limit, the MONDian potential in
TeVeS can be expressed as the combination of Newtonian potential
and a scalar field~\citep{TeVeS},
\begin{equation}\label{corrV}
  \Phi=\Xi\Phi_N+\phi\,,
\end{equation}
where $\Xi$ is a parameter in TeVeS and is approximately $1$, and
$\phi$ represents the strength of a scalar field. Moreover, the
scalar field itself is linked to the Newtonian potential via a
free function $\mu$~\citep{TeVeS},
\begin{equation}\label{2phis}
  \bm\nabla\phi=(k/4\pi\mu)\bm\nabla\Phi_N\,.
\end{equation}
where $\mu$ is a function of $|\bm\nabla\phi|$. This free function
$\mu$ should be chosen carefully in order to reproduce Newtonian
or MOND behavior at quasi-static limits.
In fact, Eq.~(\ref{corrV}), Eq.~(\ref{2phis}), and
Eq.~(\ref{MOND}) give the  relation between $\mu$ and $\tilde\mu$
of the modified Poisson equation or Milgraom's law
Eq.~(\ref{MOND})
\begin{equation}\label{tildemu}
  {1\over\tilde\mu}=\Xi+{k\over 4\pi\mu}\,.
\end{equation}
TeVeS has only two parameters ($k$ and $\Xi$) and one free
function ($\mu$). Therefore, the MONDian behavior controlled by
$\tilde\mu$ in Milgrom's law Eq.~(\ref{MOND}) could be understood
via the parameters and free function in TeVeS. On the other hand,
we are able to express the modified gravity $\bm\nabla\Phi$ (even
in TeVeS) as a function of Newtonian gravity $\bm\nabla\Phi_N$ via
$\tilde\mu$. In the following, we are going to discuss three
commonly used interpolation functions, $\tilde\mu$.

In Bekenstein's TeVeS paper~\citep{TeVeS} he proposed the
following interpolation function
\begin{equation}\label{Bekenform}
  \tilde{\mu}(x)={-1+\sqrt{1+4x\,}\over 1+\sqrt{1+4x\,}}\,,
\end{equation}
where $x\equiv|\mathbf{a}|/\mathfrak{a}_0$. The Bekenstein's form
above fails to  fit the rotation curves of spiral galaxies
~\citep{fb,Fam07}. In order to apply the MONDian lens equations to
observational data, we also study the simple form and the standard
form which fit the rotation curves better. The simple form
is~\citep{fb}
\begin{equation}\label{simpleform}
  \tilde\mu(x)={x\over 1+x}\,,
\end{equation}
and the standard form is ~\citep{M1}
\begin{equation}\label{standardform}
  \tilde\mu(x)={x\over\sqrt{1+x^2\,}}\,.
\end{equation}
Even though the standard form will lead to bi-values problem, so
is supposed to be unphysical under the framework of
TeVeS~\citep{zhfam06,angus06}, here we still compare the standard
form with the other two by treating it as empirical function and
for the sake of curiosity.

In fact, all these three forms can be included in the following
two-parameter form
\begin{equation}\label{canonicalform}
  \tilde{\mu}(x)=\left[1-{2\over (1+\eta x^\alpha)+\sqrt{(1-\eta x^\alpha)^2+4x^\alpha\,}}
  \right]^{1/\alpha}\,.
\end{equation}
Here, $(\alpha,\eta)=(1,0)$, $(1,1)$, $(2,1)$ and $(\infty,1)$
correspond to Bekenstein form, simple form, standard form and the
naive sharp-break form $\tilde\mu=\min(1,x)$, respectively. Recall
that we have set $c=1$. We note that \citet{zhfam06} have proposed
a similar expression in which they combined Bekenstein form and
simple form. We may call Eq.~(\ref{canonicalform}) the {\it
invertible canonical interpolation function} which goes from the
naive sharp-break to the smooth Bekenstein form. The corresponding
$\mu$ can be found by Eq.~(\ref{tildemu}). The nicest thing of our
{\it invertible canonical interpolation function} is that it has a
very simple counterpart in the recent Quasi-MOND theory or its
relativistic version called Bi-metric MOND~\citep{M2009,zhfam10}.
Inverting Eq.~(\ref{MOND}) with $\tilde\mu$ given by
Eq.~(\ref{canonicalform}) gives
\begin{equation}\label{invertMOND}
  -\mathbf{a}=\bm\nabla\Phi= \nu \bm\nabla\Phi_{N}\,,
\end{equation}
where
\begin{equation}\label{canonicalMOND}
  \nu(x_N) \equiv \left[1-{\eta \over 2}+
  \sqrt{x_N^{-\alpha}+ \left({\eta \over 2}\right)^2\,} \right]^{1/\alpha}\,,
\end{equation}
and $x_N\equiv|\bm\nabla\Phi_{N}|/\mathfrak{a}_0$. This analytical
result can greatly simplify the calculations in MOND.

Note that some earlier \citep{Qin95,mort1} and recent
\citep{Ferr08} physics literature formulated Milgrom's law as
\begin{equation}\label{MOND2}
  \bm{\nabla}\Phi =
  \tilde\mu^{-1/2}(x_N)\bm\nabla\Phi_N\,.
\end{equation}
However, this formulation is {\it incorrect} except for a
sharp-break function $\tilde\mu=\min(1,a/\mathfrak{a}_0)$, so
$a=\max(a_N,\sqrt{\mathfrak{a}_0 a_N})$. At this particular
situation,
\begin{equation}
  \tilde\mu^{-1}(a/\mathfrak{a}_0)=\mathfrak{a}_0/a
  =\sqrt{\mathfrak{a}_0 a_N}=\tilde\mu^{-1/2}(a_N/\mathfrak{a}_0)\,,
\end{equation}
which in general does not hold for other choices of
$\tilde\mu(x)$.

\begin{figure}
\includegraphics{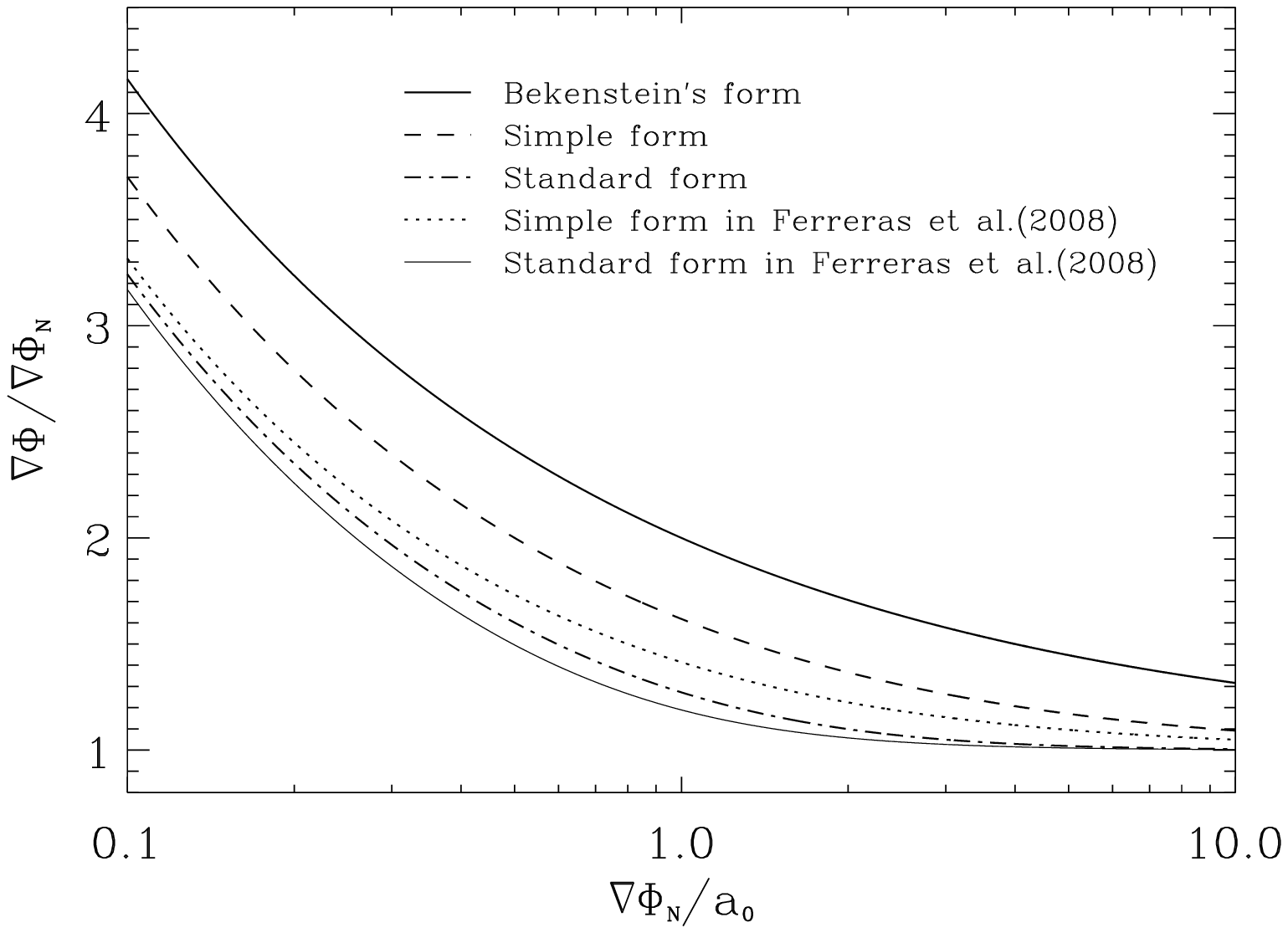}
16 \caption{\label{figure1}The strength of the MONDian gravity in
different forms of $\tilde\mu(x)$.}
\end{figure}

In Fig.~\ref{figure1}, we compare $|\bm\nabla\Phi|$ of different
forms. We also plot the results of the standard form and simple
form of the formalism adopted in~\citet{Ferr08}. For the same form
and mass, the MONDian gravity of ours is always stronger than that
of~\citet{Ferr08}.

\subsection{Doubled Image systems}

In realistic strong lensing systems, light rays often penetrate
the mass distribution of lens such that point mass model is not
appropriate.
Since MOND is supposed to have little dark matter, we adopt the
Hernquist model of elliptical galaxies~\citep{hern90} for our
lenses. The Newtonian gravity is given by
$|\bm\nabla\Phi_{N}|=GM/(\varrho+r_{h})^2$. In the following
calculations we assume $r_{h}=0.551 r_{e}$, where $r_{e}$ is the
effective radius from surface brightness
observation~\citep{hern90}.

As shown in last subsection, seeking solution
$\mathbf{a}=\bm\nabla\Phi$ of Eq.~(\ref{MOND}) depends on the
choice of $\tilde\mu$. In general we can express
$|\bm\nabla\Phi|=g(\bm\nabla\Phi_N)$, and the deflection angle
produced by a spherical lens (Eq.~(\ref{angle})) can be written as
\begin{equation}\label{angle2}
  \Delta\varphi={4GM \over\varrho_{0}} f\,,
\end{equation}
where
\begin{equation}\label{angledeflect}
  f\equiv \int^{\infty}_{\varrho_0}
  {g(\bm\nabla\Phi_N)\varrho_{0}^2\over GM (\varrho^2-\varrho_{0}^2)^{1/2}}\ d\varrho\,.
\end{equation}
Moreover, since what we can measure in sky are not deflection
angles but positions of the projected images, it is useful to
define $\theta=\varrho_0/D_L$
and $\theta_E=\sqrt{4GM D_{LS}/D_L D_S\,}$, where $D_L$,$D_S$ and
$D_LS$ are distances from the observer to lens, observer to source
and lens to source, respectively. Here $\theta_E$ is called the
Einstein radius. For a spherical strong lens, two images are
located on both sides of the lens ($\theta_{+}$ and $\theta_{-}$).
The corresponding lens equations are \citep[cf.,][]{ckt06}
\begin{eqnarray}
  \beta=\theta_{+}-{D_{LS}\over D_S}\Delta\varphi(\theta_{+})
  =\theta_{+}-{\theta_E^2\over\theta_{+}}f_{+}\,,
  \label{lenequ}\\
  \beta={D_{LS}\over D_S}\Delta\varphi(\theta_{-})-\theta_{-}
  ={\theta_E^2\over\theta_{-}}f_{-}-\theta_{-}\,,
  \label{lenequ2}
\end{eqnarray}
where $\beta$ is the source position. These two lens equations can
be combined into
\begin{equation}\label{leneq_key}
  \theta_{E}^2={\theta_{+}\theta_{-}(\theta_{+}+\theta_{-})
  \over(\theta_{+}f_{-}+\theta_{-}f_{+})}\,.
\end{equation}
Therefore, with the observed positions of the two images,
$\theta_{+}$ and $\theta_{-}$, we are able to infer the total mass
of the lens by computing $\theta_E$.


\section{Results}\label{deflectionAngle}

We apply our formalism
to 10 double-image lenses from the CASTLES Catalogue
\citep{Ru003},
of which the stellar masses have been estimated by stellar
population synthesis model with different initial mass functions
(IMFs)~\citep{Ferr05}. Table~1 lists our result in conventional
$\Lambda$CDM cosmology ($\Omega_{m}=0.3$, $\Omega_{\Lambda}=0.7$,
$h=0.7$). The lensing mass $M_L$ estimated from the Bekenstein's
form is the smallest, then the simple form and then the standard
form, on top of them is the standard form by ~\citet{Ferr08}.
Furthermore, comparison of $M_L$ from GR and TeVeS supports the
idea that the mass discrepancy between the Newtonian and the
MONDian paradigm can be quite significant.

\begin{table}
\caption{\label{tab:example}Total mass of lenses ($10^{10}
M_{\odot}$) in $\Lambda$CDM}
\begin{ruledtabular}
\begin{tabular}{rrrrrrrr}
 & \multicolumn{4}{c}{TeVeS} &  & \multicolumn{2}{c}{GR}  \\
\cline{2-5} \cline{7-8} \\
{Lens} & {Bekenstein} & {Simple}& {Standard} &
{FSY08\footnote{Ferreras et al.(2008)}} & & & {FSY08} \\
\hline
Q$0142-100$   &  19.39  &24.32& 28.34 & 29.28 && 32.29 & 32.37   \\
HS$0818+1227$ &  29.59 &37.64& 44.68 & 46.31&& 50.80 & 50.99   \\
FBQ$0951+2635$ &  2.20 & 2.72 & 3.04& 3.82&& 3.30 & 4.07    \\
BRI$0952-0115$ &  2.59 &3.35& 4.05 &  6.62&& 4.60 & 7.33    \\
Q$1017-207$ &  6.41 &8.17& 9.69 & 9.04&& 10.95 & 9.93     \\
HE$1104 - 1805$ &  64.53 & 82.99& 99.76 & 103.17 && 112.67 & 112.93   \\
LBQ$1009-025$ &  11.65 & 14.65& 17.02 &  && 19.28 &     \\
B$1030+071$ &  17.55 & 21.54& 24.40 &  && 27.08 &    \\
SBS$1520+530$ &  17.92 & 22.68& 26.51 &  && 29.57 &    \\
HE$2149-274$ &  14.33 & 18.14& 21.36 &  && 24.14 &    \\
\end{tabular}
\end{ruledtabular}
\end{table}

\begin{table}
\caption{\label{tab:2}Aperture mass (and total mass) of lenses
($10^{10} M_{\odot}$) in $\nu$HDM}
\begin{ruledtabular}
\begin{tabular}{rrrrrrrrr}
   &  & \multicolumn{3}{c}{TeVeS} &  &
\multicolumn{2}{c}{IMF (FSW05\footnote{Ferreras et al.(2005)})} \\
\cline{3-5} \cline{7-8} \\
{Lens} & {$|\nabla\Phi_{s}|/\mathfrak{a}_0$} & {Bekenstein} &
{Simple} & {Standard} &
 & {Chabrier} & {Salpeter } \\
\hline Q$0142-100$ &5.85 & 10.79 (18.36) & 13.66 (23.20) &
16.05(27.31) &&$20.9^{30.8}_{ 13.0}$  & $18.3^{32.2}_{13.2}$  \\
HS$0818+1227$ &5.67 &  18.14 (28.30) & 23.39 (36.17) & 27.79
(43.35) &&
$16.2^{21.2}_{12.6}$   & $20.8^{28.1}_{13.4}$  \\
FBQ$0951+2635$ &10.23 & 1.54 (2.16)  & 1.91  (2.67)  & 2.15 (3.01)
&&
$1.1^{2.1}_{0.5}$      & $1.5^{3.0}_{0.8}$   \\
BRI$0952-0115$ &5.28 & 2.01 (2.48)  & 2.59  (3.21)  & 3.19 (3.93)
&&
$3.5^{4.0}_{2.7}$      & $4.4^{5.2}_{3.5}$   \\
Q$1017-207$    &5.16 & 2.45 (5.89)  & 3.22  (7.55)  & 3.81 (9.15)
&&
$4.3^{13.0}_{1.4}$     & $6.4^{19.0}_{2.3}$   \\
HE$1104 - 1805$ &4.98 & 45.17 (59.58) & 58.44 (77.08) & 71.78
(94.68) &&
$22.8^{51.2}_{12.7}$   & $36.6^{63.7}_{23.1}$  \\
LBQ$1009-025$  &5.40 & 7.71 (10.79)  & 9.76 (13.67)  & 11.53
(16.15) &&
$5.5^{7.9}_{4.2}$ & $ 7.4^{9.8}_{5.0}$ \\
B$1030+071$    &8.06 & 9.76 (16.61)  & 12.06 (20.51)  & 13.80
(23.47) &&
$10.6^{15.3}_{6.5}$ & $ 14.5^{21.3}_{8.3}$ \\
SBS$1520+530$  &6.03 & 11.91 (16.67)  & 15.20 (21.28)  & 18.08
(25.31) &&
$18.5^{30.9}_{11.2}$& $21.8^{34.1}_{11.9}$ \\
HE$2149-274$   &5.86 & 7.04 (13.58)  & 8.96 (17.28)  & 10.67
(20.58) &&
$4.6^{6.7}_{3.6}$ & $6.9^{8.9}_{5.0}$ \\
\end{tabular}
\end{ruledtabular}
\end{table}

\begin{figure}[t]
\includegraphics{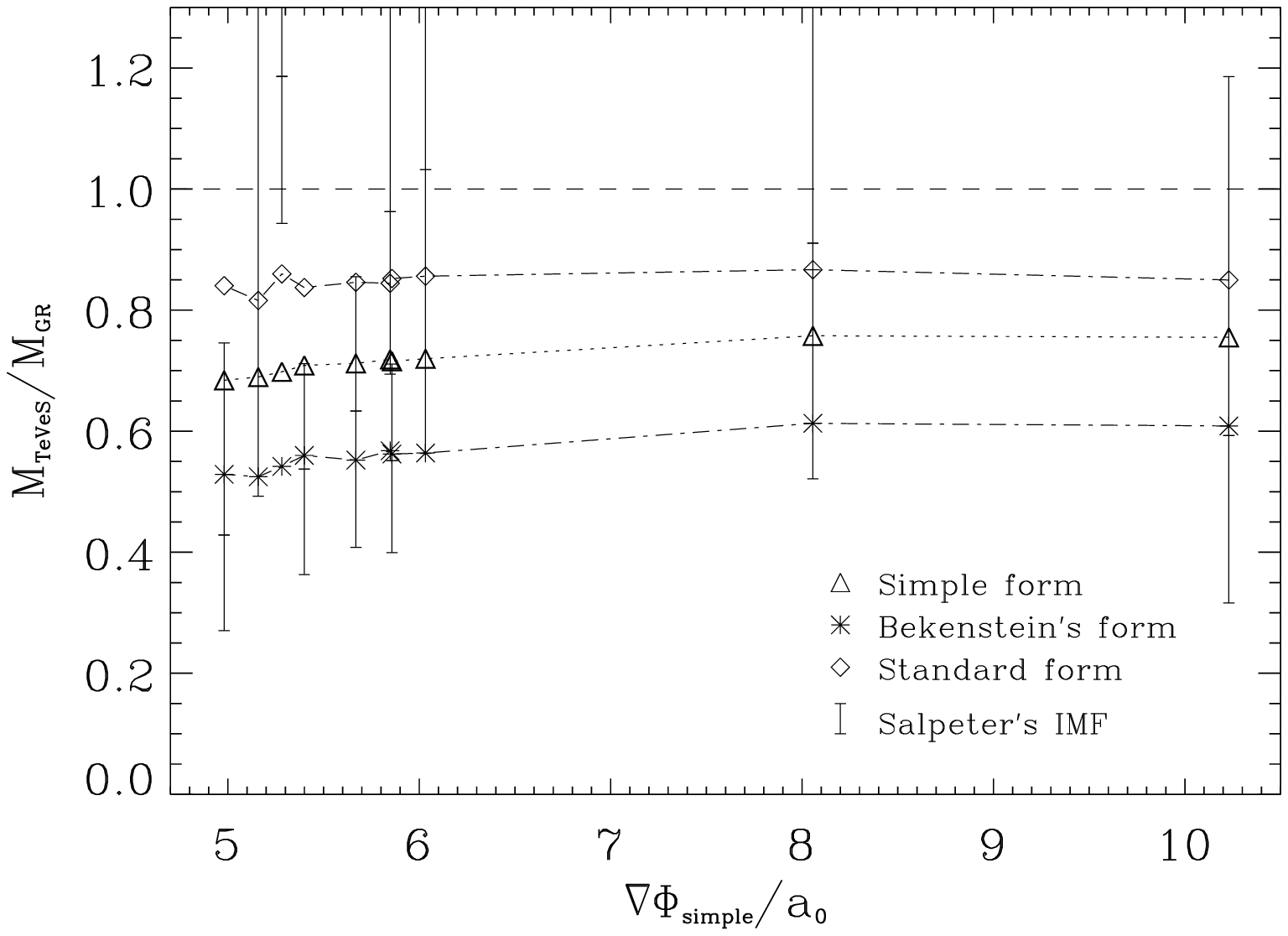}
16 \caption{\label{figure2}Mass differences between $M_{\rm GR}$
and $M_{\rm TeVeS}$ in three different forms of $\tilde{\mu}(x)$.
Estimated stellar mass using Salpeter's IMFs~\citep{Ferr05} is
shown for comparison.}
\end{figure}

\begin{figure}[t]
\includegraphics{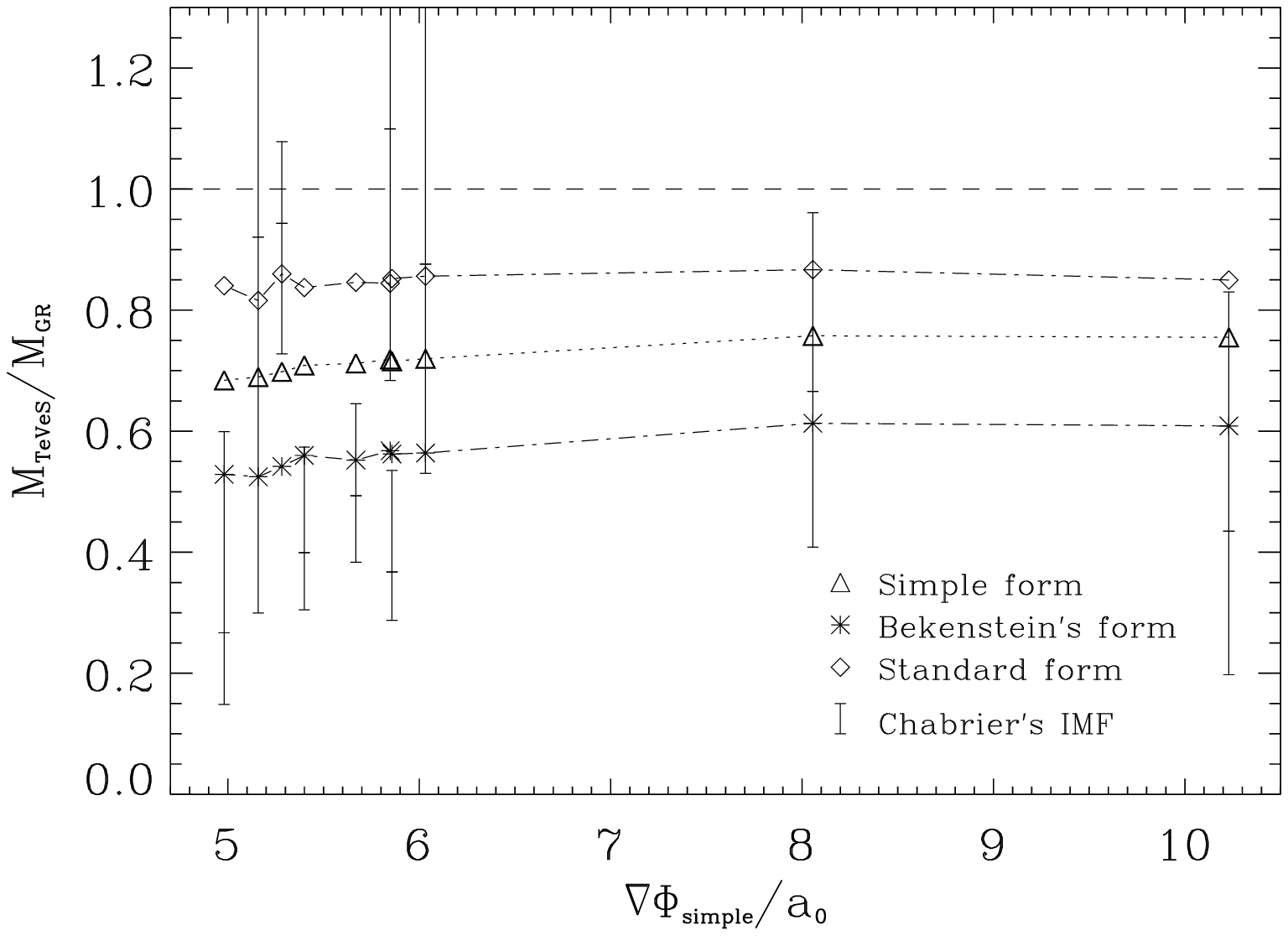}
16 \caption{\label{figure3}Same as Fig.~\ref{figure2}, except that
the estimated stellar mass used Chabrier's IMFs~\citep{Ferr05}.}
\end{figure}

For a more consistent analysis, lensing in TeVeS should be studied
in the MONDian cosmology - $\nu$HDM ($\Omega_{b}=0.05$,
$\Omega_{\nu}=0.17$ $\Omega_{\Lambda}=0.78$,
$h=0.7$)~\citep{SMFB06}.
We list the total mass of the 10 systems for the three forms of
$\tilde\mu(x)$ in the bracket of columns~3 to 5 (in bracket) of
Table~2. It appears that in TeVeS, lensing mass in the $\nu$HDM
cosmology is about $5\%$ less massive than in $\Lambda$CDM.

\citet{Ferr05} estimated the aperture stellar mass by two IMFs:
\citet{chab03} and \citet{salp55}, and the results are listed in
columns~6 \& 7 of Table~2, respectively. For comparison, we
compute the mass enclosed inside the truncated radius given in
\citet{Ferr05} from the lens total mass. The result is listed in
columns~3 to 5.

When comparing with the stellar mass ($M_{\rm Salp}$) from
Salpeter's IMF, we find that, except for BRI$0952-0115$ (where
lensing gives smaller mass), all
masses derived from the simple form are within the uncertainty of
$M_{\rm Salp}$. However, when comparing with stellar mass ($M_{\rm
Chab}$) from Chabrier IMF, Bekenstein's form is a better choice: 8
objects agree with $M_{\rm Chab}$.

Figs.~\ref{figure2} \& \ref{figure3} show the mass ratio of TeVeS
to GR for the 10 lensing systems. For comparison of mass
differences between the three forms, all ratios are plotted
against $|\bm\nabla\Phi|/\mathfrak{a}_0$ of the simple form.
We found the ratio increases slightly with
$|\bm\nabla\Phi|/\mathfrak{a}_0$. This does make sense because a
smaller $|\bm\nabla\Phi|/\mathfrak{a}_0$ means the system is
closer to the MOND regime, and a larger mass discrepancy is
expected.

\section{Discussion}\label{discussion}
In our analysis, the simple form of $\tilde\mu(x)$ (i.e.,
$\alpha=1$, $\eta=1$ in the canonical form~(\ref{canonicalform}))
yields the most reasonable lensing mass in TeVeS: in 9 of 10
systems the lens mass $M_{L}$ is within the uncertainty of $M_{\rm
Salp}$, the mass estimated from population syntthesis model with
Salpeter's IMF.
When compare with result estimated from Chabrier's IMF, 4 systems
are outside the uncertainty of $M_{\rm Chab}$, but the mass
difference is within $17^{+8.22}_{-6.64}\%$.
Moreover, the simple form yields an average of $28\%$ mass
discrepancy between the inferred lensing mass of TeVeS and GR.
This is close to the value $30\%$ estimated by ~\citet{rhs08}. For
comparison, Bekenstein's form gives an average of $44\%$, and
provides a better fit to lower-mass IMFs. This consists with the
conclusion of~\citet{nap}.

We have to keep in mind that the uncertainty in estimating the
mass of elliptical galaxies from IMF is still quite large. It is
far from mature to claim that the MONDian paradigm favors which
choice of $\tilde{\mu}(x)$ in elliptical galaxies. However, our
study shows that like spiral galaxies, elliptical galaxies can be
used to distinguish different forms of $\tilde{\mu}(x)$ as well.
We believe that along side with an independently precise
measurement of mass of elliptical galaxies is available, strong
gravitational lensing can offer us a window to study the form of
$\tilde{\mu}(x)$. We explicitly write down a formalism for this
application in the future.


We conclude that these systems do not show any apparent need of ad
hoc dark matter in elliptical galaxies. On the other hand, with a
simple spherically symmetric lens model, TeVeS is able to fit
these system reasonably well. This differs from the conclusion
of~\citet{Ferr08} but sides with~\citet{zhaoeal06} and
~\citet{shan08}, where a constant mass-to-light ratio is assumed.

In ~\citet{Ferr08}, they concluded that MOND might have problems
in explaining galactic lensing system. Ours analysis differs from
their conclusion. The reason could be that they have applied the
formalism of Eq.(\ref{MOND2}) in their calculation. Indeed, this
might also explain why they have found a considerable discrepancy
between MOND and TeVeS~\citep{mav09}.
In \citet{mav09} and \citet{Ferr09},
the full relativistic equations with vector were solved for the
first time.
They discussed the result of two forms of $\mu$ proposed by
\citet{angus06}. Their choices of $\mu$ are identical to the
simple form and Bekenstein's form in this paper, but a gap still
exists between our result and ~\citet{mav09}. The disparity might
be due to the fact that we choose Hernquist model rather than NFW
model for the lens.

Our analysis also shows that the simple form seems to give the
most reasonable angle of deflection in strong lensing, which
coincides with the studies on the dynamics of spiral
galaxies~\citep{fb,zhfam06,Fam07}. It is quite reasonable because
the deviation of TeVeS from GR in strong lensing is only due to
the change of gravity (Possion equation). So we do expect a
consistent conclusion between gravitational lensing and dynamical
analysis.

We should point out that our analysis is based heavily on the
assumption that the lenses in these systems are spherical and
quasi-static, so that we can use Eq.(\ref{weakspherical}) to
derive the lensing equation. However, the assumption of spherical
distribution is obviously insufficient for most known lenses.
Moreover, the choice of mass model for lens may also affect the
inference of the total mass of the lens. Thus, it is worthwhile to
study how the deviation from this assumption would affect our
conclusion here. We note that while a spherical Hernquist model is
a reasonable assumption for some lenses, it is always a poor
assumption for modeling rotation curves of spiral galaxies which
have often exponential profiles in their disks, and the gravity is
enhanced in the disk plane. In this regard the recent claim of
inconsistency of lensing and galaxy rotation curve in MOND
\citep{Ferr09} has yet to be corrected for these systematic
effects. Systems in cluster environment which are not yet in full
dynamical equilibrium it might be important to include non-trivial
effects of the vector field \citep{FS09,Zhao07}.

\begin{acknowledgments}
MCC is partly supported by SUPA(UK), and would like to thank Andy
Taylor for helpful discussion. CMK and YT are supported in part by
the Taiwan National Science Council grants NSC
96-2112-M-008-014-MY3 and NSC 99-2112-M-008-015-MY3. HZ
acknowledges numerous hospitalities at university of Edinburgh and
thanks especially Andy Taylor for discussions.
\end{acknowledgments}


\begin{thebibliography}{}

\bibitem[28Tegmark et al.(2006)]{teg06}M. Tegmark et al., \prd {\bf{74}}, 123507 (2006)
\bibitem[02Bekenstein(2004)]{TeVeS} J.D. Bekenstein, \prd {\bf{70}}, 083509 (2004)
\bibitem[17Milgrom(1983)]{M1}Milgrom, M. 1983, \apj, {\bf{270}}, 365
\bibitem[18Milgrom(2009)]{M2009}M. Milgrom,  \prd {\bf{80}}, 123536 (2009)
\bibitem[33Zhao \& Li(2010)]{zl10}H.S. Zhao,\&  B. Li, {\bf{712}}, 130 (2010)
\bibitem[07Dodelson \& Liguori (2006)]{dl06}S. Dodelson, \&  M. Liguori, \prl {\bf{97}}, 231301 (2006)
\bibitem[27Skordis et al.(2006)]{SMFB06}C. Skordis, D.F. Mota, P.G. Ferreira  \& C. Boehm, \prl {\bf{96}}, 011301 (2006)
\bibitem[25Sanders \& McGaugh(2002)]{SandersMcGaugh} R.H. Sanders \& S.S. McGaugh, Ann. Rev. Astron. Astrophys. {\bf{40}},  263 (2002)  
\bibitem[09Famaey et al.(2007)]{Fam07}B. Famaey, G. Gentile, J.-P. Bruneton, \& H.S. Zhao, \prd {\bf{75}}, 063002 (2007)
\bibitem[06Clowe et al.(2006)]{clowe06} D. Clowe, A. Gonzalez, \&  M. Markevitch, \apj {\bf{648}}, L109 (2006)
\bibitem[21Qin et al.(1995)]{Qin95}B. Qin, X.P. Wu,\& Z.L. Zou,  1995, A\&A {\bf{296}}, 264 (1995)
\bibitem[19Mortlock \& Turner(2001)]{mort1} D.J. Mortlock \&  E.L. Turner, Month. Not. Roy. Astron. Soc. {\bf{327}}, 557 (2001)
\bibitem[04Chiu et al.(2006)]{ckt06} M.C. Chiu, C.M. Ko,  \&  Y. Tian, \apj {\bf{636}}, 565 (2006)
\bibitem[34Zhao et al.(2006)]{zhaoeal06}H.S. Zhao, D.J. Bacon, A.N. Taylor,  \& K. Horne, Month. Not. Roy. Astron. Soc., {\bf{368}}, 171 (2006)
\bibitem[10Feix et al.(2008)]{Feix08}  M. Feix, C. Fedeli, \& M. Bartelmann, A\&A, {\bf{480}}, 315 (2008)
\bibitem[26Shan et al.(2008)]{shan08}H.S. Shan, M. Feix, B. Famaey \& H.S. Zhao, Month. Not. Roy. Astron. Soc., {\bf{387}}, 1303 (2008)
\bibitem[29Xu et al.(2008)]{xu08}D. Xu et al., \apj {\bf{682}},711 (2008)    
\bibitem[05Chiu et al.(2008)]{ckt08} M.C. Chiu,  Y. Tian, \& C.M. Ko, arXiv: astro-ph/08125011 (2008)    
\bibitem[13Ferreras et al.(2008)]{Ferr08}I. Ferreras, M. Sakellariadou,\&  M.F. Yusaf, \prl {\bf{100}}, 031302 (2008)
\bibitem[14Ferreras et al.(2009)]{Ferr09}I. Ferreras, N.E. Mavromatos, M. Sakellariadou,\& M.F. Yusaf, \prd {\bf{80}}, 103506 (2009)
\bibitem[12Ferreras et al.(2005)]{Ferr05}I. Ferreras, P. Saha,\& L. Williams, \apj {\bf{623}}, L5 (2005)    
\bibitem[08Famaey \& Binney(2005)]{fb}B. Famaey,  \& J. Binney, Month. Not. Roy. Astron. Soc. {\bf{363}}, 603, (2005)
\bibitem[01Angus et al.(2006)]{angus06}G. W. Angus, B. Famaey, \& H.S. Zhao, Month. Not. Roy. Astron.
Soc. {\bf{371}},138 (2006)
\bibitem[31Zhao \& Famaey(2006)]{zhfam06}H.S. Zhao, \& B. Famaey, \apj, {\bf{638}}, L9 (2006)
\bibitem[32Zhao \& Famaey(2010)]{zhfam10}H.S. Zhao, \& B. Famaey, \prd {\bf{81}} ,087304(2010)
\bibitem[15Hernquist(1990)]{hern90}L. Hernquist, \apj {\bf{356}}, 359 (1990) 
\bibitem[22Rusin et al.(2003)]{Ru003}D. Rusin et al., \apj {\bf{587}}, 143 (2003)   
\bibitem[03Chabrier(2003)]{chab03}G. Chabrier, PASP {\bf{115}}, 763 (2003)    
\bibitem[23Salpeter(1955)]{salp55}E.E. Salpeter, \apj {\bf{121}}, 161 (1955)
\bibitem[24Sanders(2008)]{rhs08}R.H. Sanders,  Month. Not. Roy. Astron. Soc. {\bf{389}}, 701 (2008)
\bibitem[20Napolitano et al. (2010)]{nap}  N.R. Napolitano in private communication
\bibitem[16Mavromatos et al.(2009)]{mav09} N.E. Mavromatos,  M. Sakellariadou, \& M.F. Yusaf, \prd {\bf{79}}, 081301 (2009)
\bibitem[11Ferreira \& Starkman (2009)]{FS09} I. Ferreira \& G. Starkman, Science {\bf{326}}, 812 (2009)
\bibitem[30Zhao(2007)]{Zhao07}H. Zhao, \apj, {\bf{671}}, L1 (2007)

























\end{thebibliography}
{}

\end{document}